\documentclass[twocolumn,english,aps,prb]{revtex4}
\usepackage[T1]{fontenc}
\usepackage[latin9]{inputenc}
\usepackage{bm}
\usepackage{amsmath}
\usepackage{amssymb}
\usepackage{graphicx}
\usepackage{babel}
\renewcommand{\vec}[1]{\boldsymbol{#1}}
\makeatother

\usepackage{babel}
\begin{document}

\title{Phase diagram of the spin-1 Heisenberg model with three-site interactions on the square lattice}

\author{F. Michaud$^{1}$, and F. Mila$^{1}$}

\affiliation{$^{1}$ Institute of Theoretical Physics, Ecole Polytechnique F\'ed\'erale de Lausanne (EPFL), CH-1015 Lausanne, Switzerland}

\date{\today}
\begin{abstract}
We study the spin $S=1$ antiferromagnetic Heisenberg model on the square lattice with, in addition to the nearest-neighbor interaction, a  three-site interaction of the form $\left(\vec S_{i}\cdot{\vec S_{j}}\right)\left(\vec S_{j}\cdot{\vec S_{k}}\right) + \text{h.c.}$. This interaction appears naturally in a strong coupling exansion of the two-orbital, half-filled Hubbard model. For spin 1/2, this model reduces to a Heisenberg model with bilinear interactions up to third neighbors, with a second-neighbor interaction twice as large the third-neighbor one, a very frustrated
model with an infinite family of helical classical ground states in a large parameter range. Using a variety of analytical and numerical methods, we show that the spin-1 case is also very frustrated, and that its phase diagram is even richer, with possibly the succession of seven different phases as a function of the ratio of the three-site interaction to the bilinear one. The phases are either purely magnetic phases with collinear order, or of mixed magnetic and quadrupolar character with helical order.
\end{abstract}
\maketitle

\section{Introduction}
\label{sec:introduction}
The quest for exotic phases in magnetic quantum systems has driven a lot of research over the past years. The appearance of new phases is often related to the inclusion of terms beyond the first neighbor Heisenberg interaction. For spin $S=1/2$, it can be for example longer range interactions leading to frustration\cite{book_FM}, Dzyaloshinskii-Moriya interactions\cite{Dzyaloshinsky,Moriya} or plaquette interactions\cite{Coldea}. These terms also appear in spin $S=1$ systems. However, the fact that the local Hilbert space is of higher dimension for $S=1$ allows for new types of interaction. Starting from a two-orbital Hubbard model at half-filling with Hund's coupling, a strong coupling expansion leads to an effective spin-1 model. At second order, only a Heisenberg interaction between first neighbor appears. At fourth order, on top of the usual terms that also appear in the $S=1/2$ case starting from the single band Hubard model\cite{Takahashi,macdonald}, namely second-neighbor and plaquette interactions, two new interactions appear, the 
biquadratic interaction, and a three-spin interaction \cite{PhysRevB.76.132412,MVMM} of the form $\left(\vec S_{j}\cdot{\vec S_{i}}\right)\left(\vec S_{i}\cdot{\vec S_{k}}\right) + \text{h.c.}$. The biquadratic interaction has been intensively studied over the past few years, both in one dimension\cite{lai,Sutherland,Takhtajan2,Babujian2,AKLT,Chubukov,Fath1,Fath2,Jolicoeur,Laeuchli,Salvatore} and two dimensions\cite{Kawashima,BLBQtsunetsugu,BLBQmila,tamas}.
However to the best of our knowledges, the effect of the three-spin interaction has not been investigated in 2D, which has led us to study the following model on a square lattice:
\begin{eqnarray}
\label{eq:ham}
H&=&J_1\sum_{<i,j>} \vec S_{i}\cdot \vec S_{j} \nonumber \\
&+&
J_3 \sum_{<i,j,k>}\left(\vec S_{i}\cdot \vec S_{j}\right) \left(\vec S_{j}\cdot \vec S_{k} \right) + \text{h.c.}
\end{eqnarray} 
where $\vec S_i$ are spin $S=1$ operators, $<i,j>$ sums over first neighbors and $<i,j,k>$ sums over all possible configurations where $i$ and $k$ are first neighbors of $j$ and are different from each other. 

It is worthwhile noticing that in the case of spin-1/2, the three-spin interaction reduces to a second-neighbor interaction. On a chain, this  leads to the $J_1-J_2$ model which has a transition to a dimerized ground state at $J_2/J_1 \approx 0.2411$\cite{J1J2transition}. In Ref. \onlinecite{MVMM}, it has been shown that the three-spin interaction generalizes this result to higher spins and that, for any $S$, the ground state shows dimerization in some region of the phase diagram. In two dimensions and for spin-1/2, the Hamiltonian of eq. (\ref{eq:ham}) reduces to a Hamiltonian with  Heisenberg interactions $J_1$,$J_{2^{\text{nd}}}$ and $ J_{3^{\text{rd}}}$ to first, second and third neighbor with $J_{2^{\text{nd}}} = 2 J_{3^{\text{rd}}}$. A discussion of this model can be found in Ref. [\onlinecite{J1J2J3rastelli,J1J2J3chandra,J1J2J3,J1J2J3ferrer,J1J2J3gochev,J1J2J3yang, J1J2J3m,J1J2J3Arlego}]. The line $J_{2^{\text{nd}}} = 2 J_{3^{\text{rd}}}$ is of particular interest because it lies at the classical transition between a helical phase with pitch vector $(Q,\pi)$ and another 
helical phase with pitch vector $(Q,Q)$. Starting from an antiferromagnetic phase with N\'eel order, the system undergoes at $J_{2^{\text{nd}}}/J_1 = 1/4$ a phase transition to a phase with an infinite number of helical ground states whose pitch vectors are defined by the condition:
\begin{eqnarray}
 \cos Q_x  + \cos Q_y = -\frac{J_1}{2J_{2^{\text{nd}}}}
\end{eqnarray}
At the level of linear spin-wave theory, quantum fluctuations have been shown to completely disorder this phase. This is the first hint that frustration might be very high in the case of the Hamiltonian of Eq. (\ref{eq:ham}) as well. As we shall see, the situation is even more complicated, and the competition between several phases leads to a very rich phase diagram. 

The paper is organized as follows. In section \ref{sec:classical}, we determine the different phases in the classical limit and in particular we discuss the degeneracy of the phase appearing in the limit $J_3\gg J_1$. In section \ref{sec:mean-field} we discuss the mean-field phase diagram based on a product of local wave-functions and the difference between this phase diagram and the classical phase diagram. We add quantum fluctuations to the system in the context of a semiclassical expension around the classical solutions in section \ref{sec:spin-wave}. To confirm the previous results, we give some exact diagonalization results in section \ref{sec:exact-diagonalisation}. We finish the paper by a conclusion in section \ref{sec:conclusion}

\section{Classical phase diagram}
\label{sec:classical}
In this section, we consider the spins as three dimensional arrows of length $S=1$. The zero temperature phase diagram is obtained by minimizing the energy of Hamiltonian (\ref{eq:ham}) under this assumption. The classical phase diagram can be formally established by decomposing the Hamiltonian into a sum of local Hamiltonians in the following way:
\begin{eqnarray*}
 H &=& \sum_{i}^{N} H_{i}\\
 H_{i} &=& \frac{1}{2} J_1\sum_{\vec{\tau}}{\vec S_{i}}\cdot {\vec S_{i+\vec{\tau}}}\\
 & + &2 J_3 \sum_{\vec{\tau}} \sum_{\vec{\tilde{\tau}}\neq \vec{\tau} } \left(\vec S_{i+\vec{\tau}}\cdot{\vec S_{i}}\right)\left(\vec S_{i}\cdot{\vec S_{i+{\vec{\tilde{\tau}}}}}\right),
\end{eqnarray*}
where $N$ is the number of sites, and $\vec{\tilde{\tau}}$ and $\vec{\tau}$ sum over the nearest neighbors. Minimizing independently all these local Hamiltonians is a sufficient, though not necessary, condition to have a global minimum. We will see that we were able to independently minimize the local Hamiltonian for all values of the ratio $J_3/J_1$.


For symmetry reasons, the central spin of the local Hamiltonian can be chosen to point in the $z$ direction. Since the energy only depends on the relative angle between the central spin and its four neighbor (and not on the angle between the neighbors themselves, as would be the case with for example a second neighbor interaction), all the spins can be assumed to point in the same plane, say the $z-x$ plane.
Under these assumptions, the classical energy can be written:
\begin{eqnarray*}
\label{eq:en_local}
 E&=& J_1/2\left( \cos \theta_1 +\cos \theta_2+\cos \theta_3+\cos \theta_4\right) \\
 &+& 2 J_3 \left( \cos \theta_1 \cos \theta_2 +\cos \theta_1 \cos \theta_3+\cos \theta_1 \cos \theta_4 \right.\\
 & + & \left. \cos \theta_2\cos \theta_3+\theta_2\cos \theta_4+\theta_3\cos \theta_4\right).
\end{eqnarray*}
where $\theta_i, i=1,...,4$ are the angles of the four neighbors of the central spin with respect to $z$.
This energy can be easily minimized, and we find three different types of minima, depending on the value of $J_3/J_1$. These three different local minima can be extended to the entire lattice and constitute the three phases that appear in the classical phase diagram. We will now describe in detail the three different phases. 

\subsection{N\'eel phase}
In the limit where  $J_3$ goes to zero, the classical ground state is of N\'eel type. The energy per site is given by $E_{\text{cl}}=-2 J_h + 12 J_3$.
\subsection{Up-up-down-down phase}
\begin{figure}[t]
\includegraphics[width=0.3\textwidth]{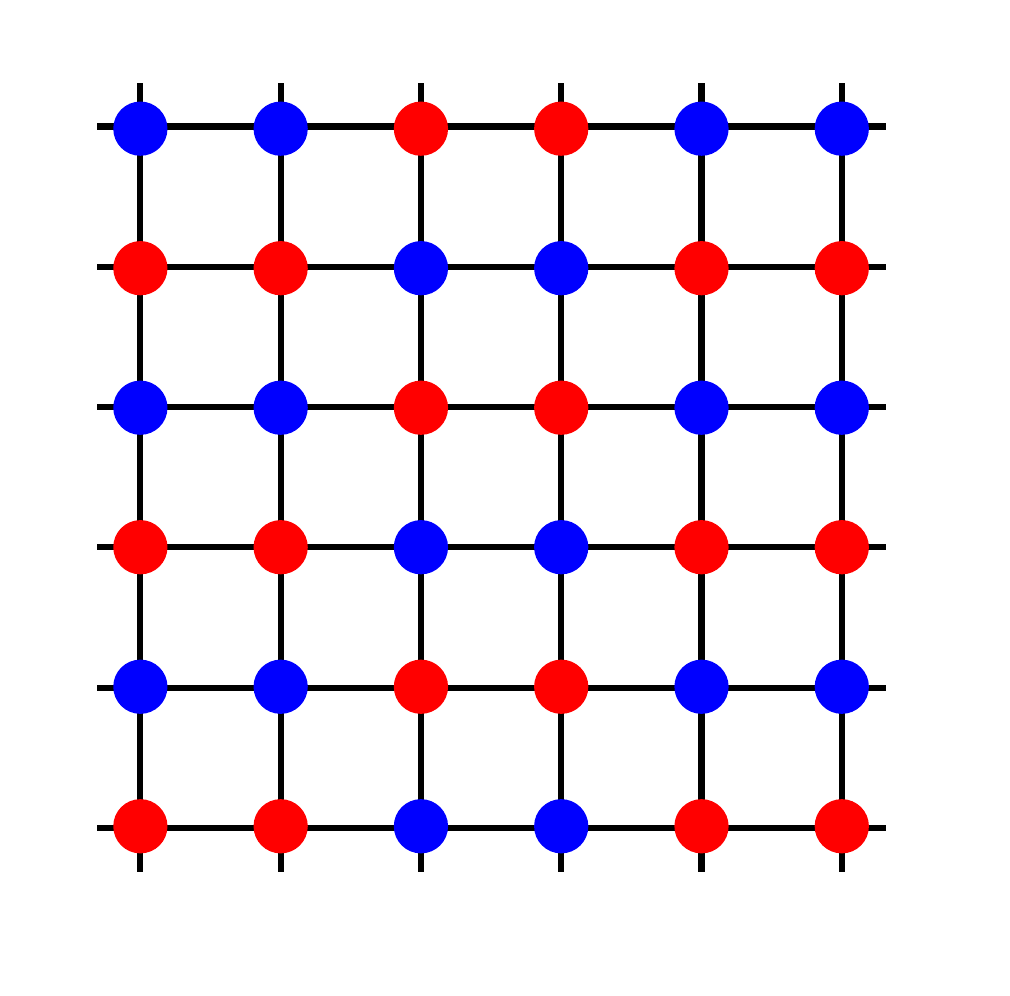}
\caption{(Color online) Classical phase for $1/12 < J_3/J_1 < 1/4 $. Blue dots mean spins up and red dots spins down. This phase is called throughout the paper the up-up-down-down (uudd) phase.}
\label{fig:intermediate}
\end{figure}

The first transition that appears when increasing $J_3$ leads to the phase depicted in Fig. \ref{fig:intermediate}. This phase realizes a compromise between the Heisenberg interaction and the three-body interaction. The three-body interaction favors phases where we have a local up-up-down or down-down-up structure. In the intermediate phase, this is realized by having up-up-down-down chains which are coupled antiferromagnetically so that the Heisenberg interaction is still partially satisfied. The energy per site is then given by $-J_1$. The transition point can be established by comparing the classical energy of the two phases. We see that the transition to this phase appears already at $J_3=J_1/12$. In the following, we will refer to this phase as the up-up-down-down (uudd) phase. 

\subsection{$J_3$ phases}

\begin{figure}[t]
\includegraphics[width=0.5\textwidth]{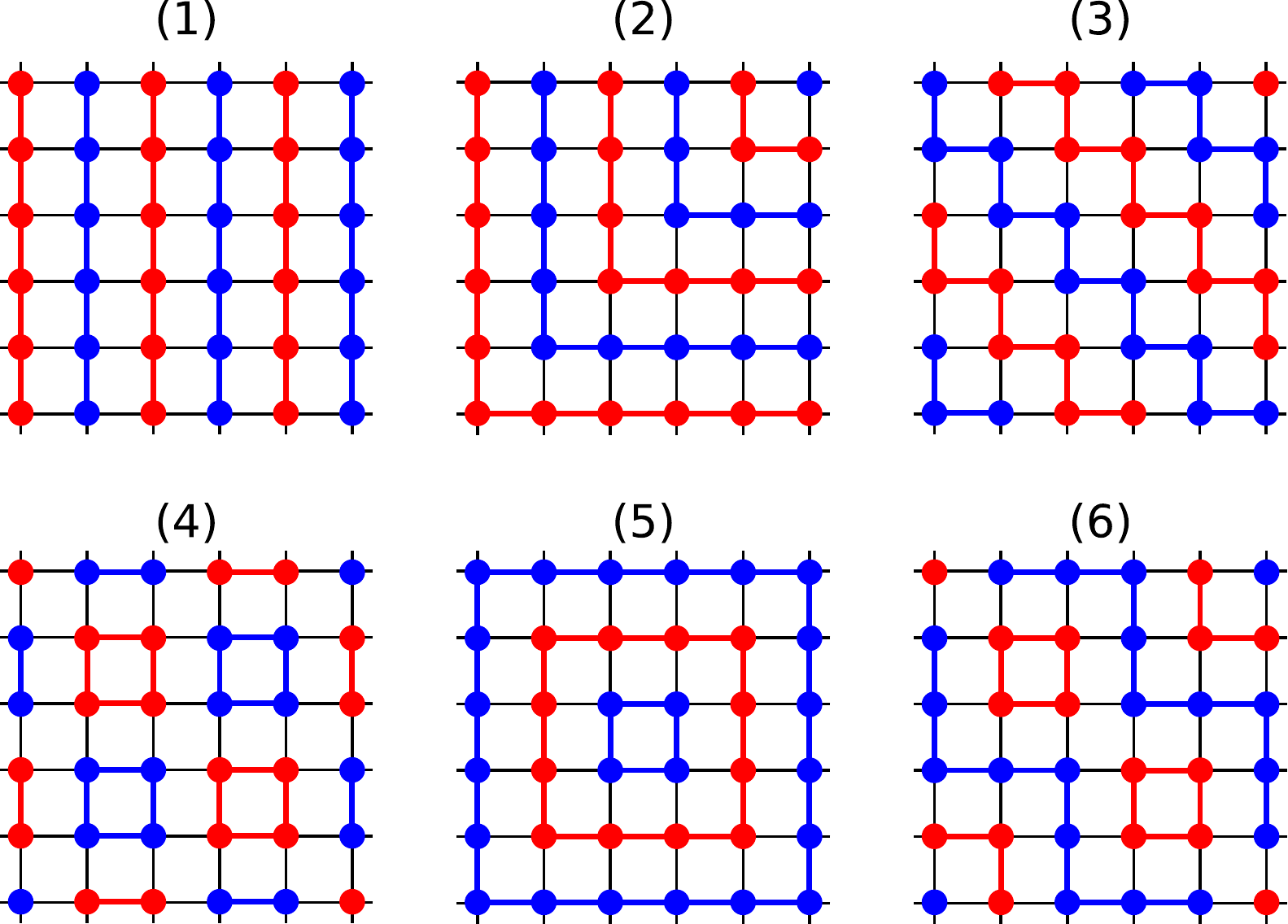}
\caption{(Color online) Some possible classical phases when $J_3/J_1 > 0.25 $. Blue dots mean spins up and red dots spins down. The color lines are guides to the eyes. }
\label{fig:phase_J3}
\end{figure}

 If $J_1=0$, the minimization of the local Hamiltonian is trivial. Because of the rotational invariance of the Hamiltonian, the central spin can point in any direction. If we choose it to point up, then the minimum is realized by having two neighbors pointing up, and two neighbors pointing down. This still holds if the central spin points down. A global minimum can be found if we manage to build a configuration where every spin has two neighbors pointing up and two neighbors pointing down. An easy way to build phases which respect this constraint is to draw lines of spins up and lines of spins down so that two lines of spins up (or spins down) never touch each other. Several examples are given in Fig. \ref{fig:phase_J3}. In the first one (1) we draw straight lines. This can be deformed by adding domain walls where all lines form an angle, as in (2). In (3), we put all possible domain walls. A second possibility shown in (4) is to draw $2\times2$ squares. This phase can also be transformed by 
adding bigger squares around the $2\times2$ square (5). Finally, we can also mix phases with closed loops and phases with lines as seen in (6).
 
\begin{figure}[t]
\includegraphics[width=0.35\textwidth]{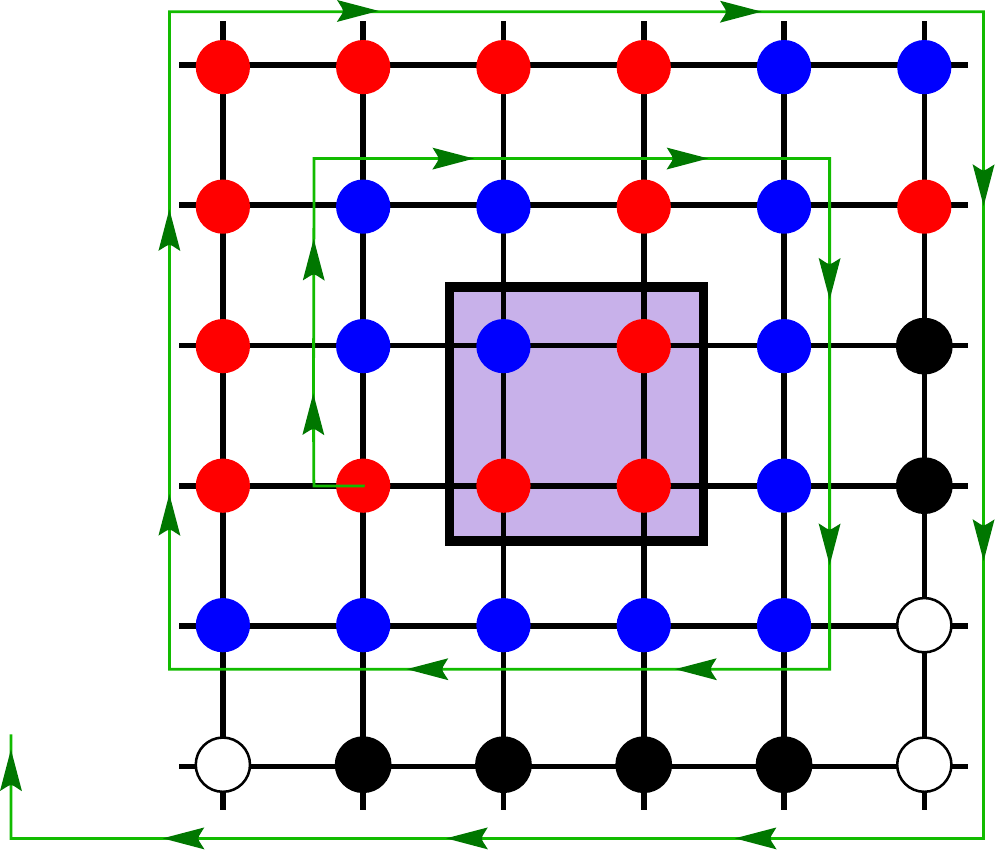}
\caption{(Color online) Construction of a ground state following a spiral. The up spins are shown in blue, the down spins in red. The black spins are completely constrained by their environment. The white ones may be constrained in some situations, but can be chosen freely in general. Since there are at most two unconstrained spins per corner, the number of free spins divided by the number of spins is going to zero in the thermodynamic limit.}
\label{fig:build_phase}
\end{figure}

 The local constraint seems weak. However, the residual entropy scales as $L$ and not as $L^2$ (where L is the linear size of the system). To prove it, let us construct a classical ground state as described in Fig.  \ref{fig:build_phase}. We start from an arbitrary cell of four spins (in purple in Fig. \ref{fig:build_phase}). Then, by turning around this cell following a spiral, we put new spins who respect the constraint imposed by the spins in the inner side of the spiral. The spin along the sides are completely constrained by the inner spins (in black on Fig. \ref{fig:build_phase}). At most two spins when the path form an angle might\footnote{We say "might" and not "can" because two neighbors are enough to constrain one spin if they are pointing in the same direction} be choosen freely (in white in Fig \ref{fig:build_phase}). The number of free spins will therefore scales at best as $8 L$ and the number of possible configuration as $2^{8L}$. This is only an upper bound for the degeneracy. Now, one can 
build at least $2^{L+1}$ ground state in a system of linear size $L$ by adding from $0$ to $L$ domain walls to the phase with straight lines of Fig. \ref{fig:phase_J3} (1) which give us a lower bound. Therefore, the residual entropy in the thermodynamic limit lies between $L$ and $8L$ and scales as the linear size of the system $L$. 
 
 For classical spins, the Heisenberg interaction does not lift the degeneracy between these different phases, since it is zero for all of them (by definition of the constraint). In the classical case, we therefore expect all these phases to coexist, even for finite $J_1$. The energy per site of this phase is $E=2J_h - 4J_3$. The transition to the intermediate phase takes place at $J_3/J_1=1/4$.

\begin{figure}[t]
\includegraphics[width=0.48\textwidth]{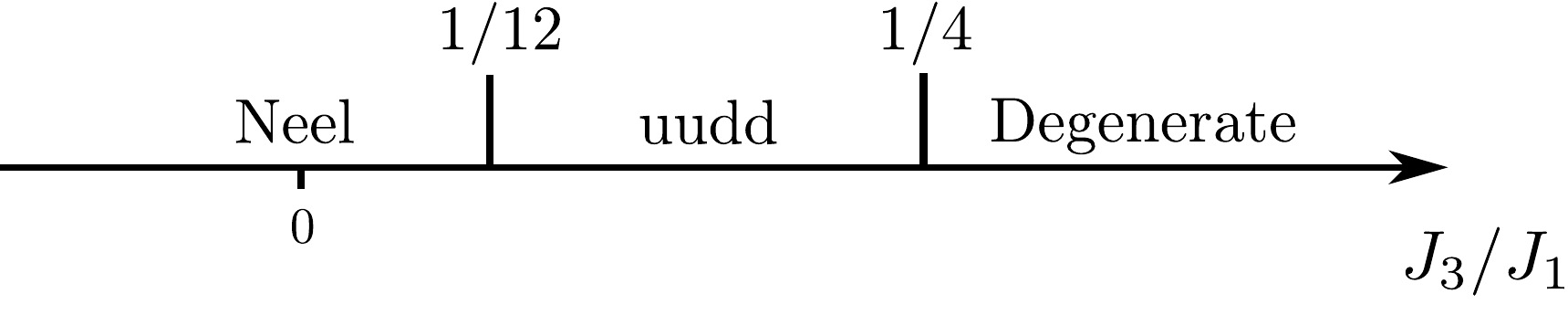}
\caption{Classical phase diagram. The phase $uudd$ can be seen in Fig. \ref{fig:intermediate}. Some examples of phases in the degenerated region can be seen in Fig. \ref{fig:phase_J3}}
\label{fig:pd_classical}
\end{figure}

Overall, the classical phase diagram contains three regions, the last one showing a big degeneracy. It is sketched in Fig. \ref{fig:pd_classical}.

\section{Mean-field phase diagram}
\label{sec:mean-field}
While classical phases give in general some first intuition about what the quantum phase diagram could be, they still show some down sides. One of them is the fact that if one considers quantum spin $S=1$ systems, the local order parameter can be of quadrupolar type instead of magnetic type. This kind of local order is known to be of crucial importance when biquadratic interactions enter the game\cite{intro_spin_nematic}. Since the three-body interaction involves the square of some local operator, it is legitimate to wonder if, in this case as well, the quadrupolar component plays a role. To tackle this difficulty, it is useful to do a mean-field phase diagram instead of a classical phase diagram\cite{intro_spin_nematic}.

To be more precise, the mean-field approach consists in reducing the Hilbert space to states which are products of local wave functions, i.e. to consider only states of the form:
\begin{eqnarray*}
|\Psi\rangle = \otimes_i |\psi_i\rangle
\end{eqnarray*}
 The local wave function for a spin $S=1$ is in general described by three complex numbers. However, the condition that the norm is one, and the freedom to fix the phase, reduce the freedom to only four real parameters. The local wave function can be chosen as:
\begin{eqnarray*}
 |\psi_i\rangle= e^{i\gamma_{i}} \cos \theta_i \cos \phi_i |1\rangle + e^{i\tilde{\gamma_{i}}}\cos \theta_i \sin \phi_i |\bar{1}\rangle + \sin \theta_i  |0\rangle
\end{eqnarray*}
where $\gamma_{i}$, $\tilde{\gamma_{i}}$, $\phi_i$ and $\theta_i$ are real numbers. It is easy to check that the norm of the local wave function is one and therefore $\langle \Psi | \Psi \rangle = 1$. The mean-field approach consists then in minimizing the mean-field energy $E = \langle \Psi | H | \Psi \rangle$, which is a function of $4 n$ variables.

One of the main reasons to use this approach is to detect quadrupolar phases, as shown in the case of biquadratic interactions\cite{BLBQmila,tamas}. However,
even for purely magnetic states, the phase diagram can be different from the classical one if one considers terms beyond the Heisenberg interaction. This effect can be traced back to the fact that the mean-value of a product of operators is the product of the mean-values of these operators only if these operators commute with each other. Since the spin operators on different sites commute, for a Hamiltonian which is linear in spin operator on all sites, the classical and the mean field values are the same. For example, for a Heisenberg Hamiltonian:
\begin{eqnarray*}
E_{\text{M.-F.}}&=&\langle \psi | H | \psi \rangle = \sum_{<i,j>}\langle \psi |{\bf S}_i{\bf S}_j| \psi \rangle
\\&=& \sum_{<i,j>}\langle \psi |{\bf S}_i| \psi \rangle \langle \psi |{\bf S}_j| \psi \rangle = E_{\text{classic}}
\end{eqnarray*}

By contrast, if one considers a Hamiltonian with higher order on-site terms such as $S_i^{\alpha}\cdot S_i^{\beta}$, which appear in the biquadratic and in the three body interactions, the decomposition as a product of mean field terms cannot be performed and the classical and mean-field energies are different, even for coherent states. As an example, let us compare the mean field and the classical energy of the biquadratic interaction in the $|1\bar{1}\rangle$ state:
\begin{eqnarray*}
E_{\text{classic}} &=& \left( 
\langle 1\bar{1} | {\bf S_1}|1\bar{1} \rangle\langle1\bar{1}| {\bf S_2}|1\bar{1}\rangle\right)^2 \\&=& \left((0,0,1)\cdot(0,0,-1)\right)^2 =  1 \\
E_{\text{M.-F.}}&=&\langle 1\bar{1}| \left( \bf S_1 \bf S_2\right)^2|1\bar{1} \rangle = \langle 1\bar{1} | {\bf S_1 }{\bf S_2}(| 00\rangle-|1\bar{1}\rangle) = 2.
\end{eqnarray*}
The two energies are indeed different. In the $J_1-J_3$ model presented here, this difference between the classical and the mean-field energy for a given configuration plays a more important role than the possibility to have quadrupolar order, and it lies at the root of the appearance of helical phases which are not present in the classical case.

The strategy to find the phase diagram is the following. First, we minimize the mean-field energy of finite clusters, up to $N=8 \times 8 $, leaving all $8 \times 8 \times 4 = 256 $ variables free. Based on the results, we do an educated guess to reduce the number of free variables to a number which does not scale with the system size (for example the angle between nearest-neighbor sites). This allows us to find the energy in the thermodynamic limit with only a few angles left free. We then have to check that the trial wave function gives an energy per site which is smaller or equal to the energy of finite systems. However, we can never be sure that a bigger cluster would not lead to a better energy, and therefore that we are not missing some phases. 

Obviously, the classical phases are included in the Hilbert space of the mean-field phase. They can therefore appear in the phase diagram. Indeed, the three classical phases are realized in some region of the phase diagram. However, new phases appear between the classical phases. Based on the results obtained from the numerical simulations, we assume that the system has either a unit cell of $2\times 1$ or is an helical phase obtained by the rotation of a $2\times 1$ cell. We also assume that the spins lie all in the $x-y$ plane. Finally, we assume that the length of the quadrupole is the same on every site. This provides us with an energy with only four independent variables, one for the relative angle inside the unit cell, two for the rotation of the unit cell in the $x$ and $y$ directions, and a last one for the length of the spin. Using this assumption, we can find an expression for the energy in the thermodynamic limit. However, the exact form of the energy is still too complicated to be minimized by 
an 
analytical treatment. Nevertheless, it can be minimized numerically with a very high precision. In the case of a second order phase transition, we can also predict the critical value $\left(J_{3}/J_1\right)_c$ where the phase changes to a helical phase by taking the Taylor expansion of the energy around the different classical phases. The transition point is given by the condition that the quadratic term vanishes. 

\subsection{First helical phase}
Starting from the N\'eel state, we have a first instability at $J_3 = J_1/14$. This state is formed by exactly antiferromagnetic chains along one direction, while along the other direction, the angle between the spins is given by $\theta=\pi+\epsilon$. Close to the transition, $\epsilon\propto \sqrt{\frac{J_3}{J_1}-\left(\frac{J_3}{J_{1}}\right)_c}$. The quadrupolar order is not driving the transition. However, once the helical phase appears, the spins develop a small quadrupolar component. The phase ends with a first order transition to the up-up-down-down phase, around 0.087. This phase is depicted in Fig. \ref{fig:helic} (a).

\subsection{Second helical phase}
The mechanism leading to the second helical phase is similar. We start from an up-up-down-down phase, and we start to tilt the spins in the direction where they are coupled antiferromagnetically. The transition point is at $J_{3} = J_1/6$. This is also a second order phase transition, and the deviation with respect to the classical phase will also behave as $ \sqrt{\frac{J_3}{J_1}-\left(\frac{J_3}{J_{1}}\right)_c}$. As in the previous case, a small quadrupolar component appears. The phase ends with a first order transition to a third helical phase at $J_3/J_1=0.25$. This phase is depicted in Fig. \ref{fig:helic} (b).

\subsection{Third helical phase}
\begin{figure}[h]
\includegraphics[width=0.48\textwidth]{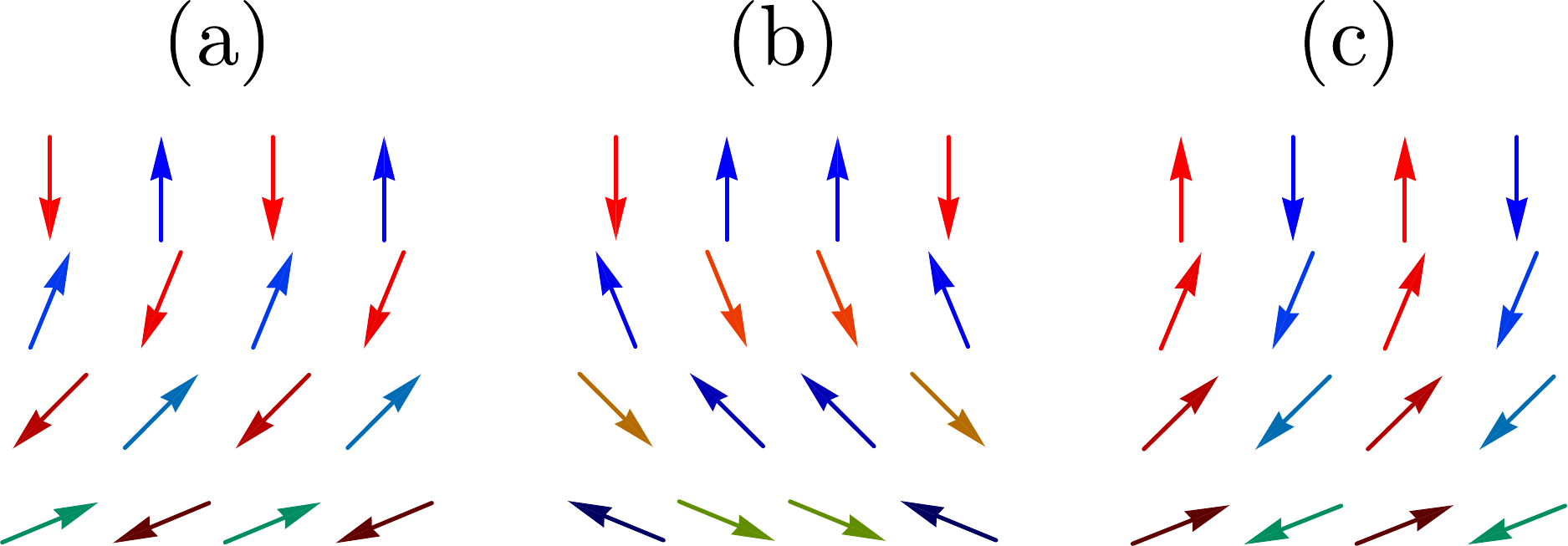}
\caption{(color online) Different helical phases which appear in the mean-field phase diagram. See text. The color are only guides to the eyes.}
\label{fig:helic}
\end{figure}

To understand the last helical phase, one has to start from the phase $(\pi,0)$ (see Fig. \ref{fig:phase_J3} (1)). The tilt starts in the ferromagnetic direction at $J_3/J_1 = 1/2$ coming from larger $J_3$, and will end at $J_3/J_1=0.25$ with the first order transition to the second helical phase. This phase is depicted in Fig. \ref{fig:helic} (c).

\subsection{Phase diagram}
\begin{figure}[t]
\includegraphics[width=0.48\textwidth]{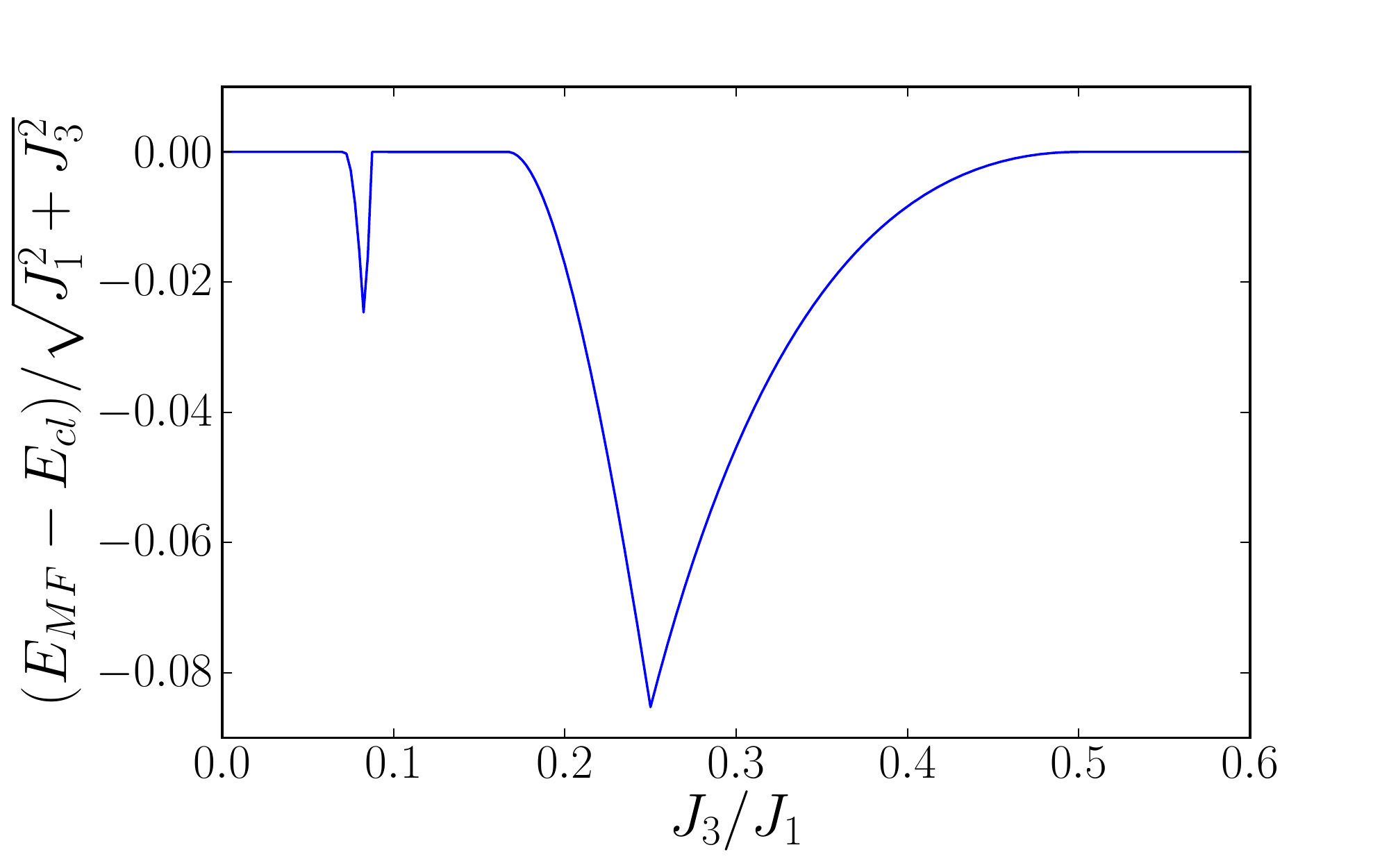}
\caption{Difference between the classical energy and the mean-field energy. Most of the time the energies are the same, but when a helical state appears, the mean-field solution become lower in energy than the classical solution.}
\label{fig:E_MF}
\end{figure}

\begin{figure}[t]
\includegraphics[width=0.48\textwidth]{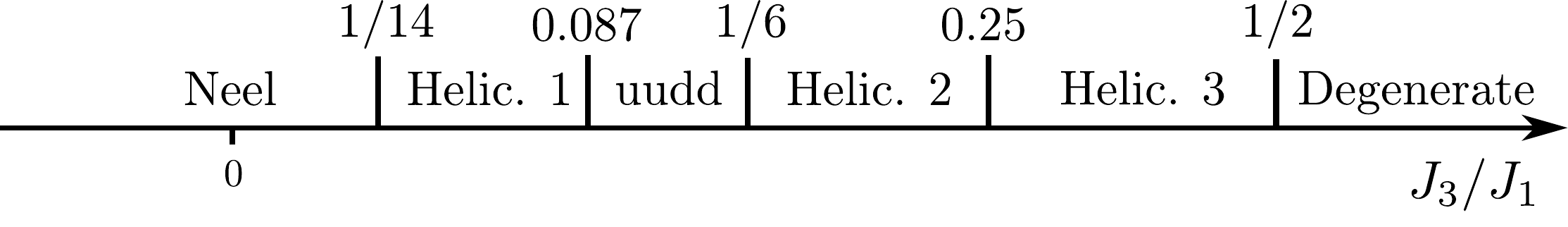}
\caption{Mean-field phase diagram. See text for a description of the phases. We can not exclude the stabilization of other phases.}
\label{fig:pd_mean_field}
\end{figure}

Overall, the three classical phases show up at some point of the mean-field phase diagram, and on top of that, three helical phases appear between the classical phases. In Fig. \ref{fig:E_MF} we show the difference between the mean-field energy and the classical energy. The mean-field phase diagram can be seen in Fig. \ref{fig:pd_mean_field}. 

\section{Quantum fluctuations}
\label{sec:spin-wave}
While the first two phases of the classical phase diagram are expected to show up in the quantum case, with of course some renormalization of the boundary, the third phase is more subtle. We expect some lifting of the degeneracy by quantum fluctuations, i.e. by a process of order by disorder. To see which phase is selected by the quantum fluctuations, we use linear spin wave theory. 

The Hamiltonian for any classical phase can be decomposed as a sum over the magnetic unit cell of local Hamiltonians:
\begin{eqnarray*}
H&=&\sum_{j}\sum_{n=1}^{N_u}\left[ \sum_{\vec{\tau}_n} \frac{J_1}{2S^2}\left(\vec S_{j,n}\cdot{\vec S_{j,n,\vec{\tau}_n}}\right) \right.\\
&+&
\left. \frac{J_3}{S^4}\sum_{\vec{\tau}_n\neq\vec{\tilde{\tau}}_n}\left(\vec S_{j,n,\vec{\tau}_n}\cdot{\vec S_{j,n}}\right)\left(\vec S_{j,n}\cdot{\vec S_{j,n,\vec{\tilde{\tau}}_n}}\right) + h.c.\right]
\end{eqnarray*} 
where $j$ runs over the different magnetic cells, $n$ runs over the elements of one unit cell, and $\vec{\tilde{\tau}}$ and $\vec{\tau}$ run over the neighbors of site $(j,n)$.

We can then perform a Holstein-Primakov transformation\cite{PhysRev.58.1098} and expand it in power of $1/S$. For a collinear phase, this leads to:
\begin{eqnarray*}
{S_x}_{j,n} &=& e^{i\vec Q \vec r_{j,n}}\sqrt{S/2}\left(a+a^{\dagger}\right)\\
{S_y}_{j,n} &=& -i\sqrt{S/2}\left(a-a^{\dagger}\right)\\
{S_z}_{j,n} &=& e^{i\vec Q \vec r_{j,n}}\left(S-n\right),\\
\end{eqnarray*} 
where $S$ is the length of the spin, $Q$ is the pitch vector of the phase, $\vec r_{j,n}$ is the position of the spin $(j,n)$, and $a$ and $a^\dagger$ represent Holstein-Primakov bosons. 

Under this transformation, keeping only the term of order $1$ (classical energy) and the term of order  $1/S$ (two-boson interactions), the Hamiltonian reads:

\begin{widetext}
\begin{eqnarray*}
H&=&\frac{J_1}{2S^2}\sum_{j}\sum_{n=1}^{N_u}\sum_{\vec{\tau}_n}\left[ e^{i\vec Q \vec \tau}\left(S^2-S\left(n_{j,n}+n_{j,n,\vec{\tau}}\right)\right)\right.\\
&+&\frac{S}{2}\left(e^{i\vec Q \vec \tau}-1\right)\left(a_{j,n}a_{j,n,\vec{\tau}}+a^\dagger_{j,n}a^\dagger_{j,n,\vec{\tau}}\right)
+\left.\frac{S}{2}\left(e^{i\vec Q \vec \tau}+1\right)\left(a_{j,n}a^\dagger_{j,n,\vec{\tau}}+a^\dagger_{j,n}a_{j,n,\vec{\tau}}\right)\right]\\
&+&\frac{J_3}{S^4}\sum_{j}\sum_{n=1}^{N_u}\sum_{\vec{\tau}_n\neq\vec{\tilde{\tau}}_n}
\left[ e^{i\vec Q (\vec \tau + \vec{\tilde{\tau}})}\left(S^4-S^3\left(n_{j,n,\vec{\tilde{\tau}}}+2 n_{j,n}+n_{j,n,\vec{\tau}}\right)\right)\right.\\
&+&\frac{S^3}{2}e^{i\vec Q \vec \tau}\left(e^{i\vec Q \vec{\tilde{\tau}}}-1\right)\left(a_{j,n}a_{j,n,\vec{\tilde{\tau}}}+a^\dagger_{j,n}a^\dagger_{j,n,\vec{\tilde{\tau}}}\right)
+\frac{S^3}{2}e^{i\vec Q \vec \tau}\left(e^{i\vec Q \vec{\tilde{\tau}}}+1\right)\left(a_{j,n}a^\dagger_{j,n,\vec{\tilde{\tau}}}+a^\dagger_{j,n}a_{j,n,\vec{\tilde{\tau}}}\right)\\
&+&\frac{S^3}{2}e^{i\vec Q \vec{\tilde{\tau}}}\left(e^{i\vec Q \vec \tau}-1\right)\left(a_{j,n,\vec{\tau}}a_{j,n}+a^\dagger_{j,n,\vec{\tau}}a^\dagger_{j,n}\right)
+\left.\frac{S^3}{2}e^{i\vec Q \vec{\tilde{\tau}}}\left(e^{i\vec Q \vec \tau}+1\right)\left(a_{j,n,\vec{\tau}}a^\dagger_{j,n}+a^\dagger_{j,n,\vec{\tau}}a_{j,n}\right)\right]\\
\end{eqnarray*} 
\end{widetext}

It is then well known that, because of the periodicity of the lattice, the Fourier transform of this expression will decouple the Hamiltonian. Taking the following convention:
\begin{eqnarray*}
 \vec a_j &=& \frac{1}{\sqrt{N}} \sum_k a_k e^{i k r_j}
 \\ \vec a_j^\dagger &=& \frac{1}{\sqrt{N}} \sum_k a_k^{\dagger} e^{-i k r_j}
\end{eqnarray*}
where $N$ is the number of sites and $r_j$ is the position of site $j$, the Hamiltonian reads:

\begin{widetext}
\begin{eqnarray*}
H&=&\frac{J_1}{4S}\sum_{k}\sum_{n=1}^{N_u}\sum_{\vec{\tau}_n}\left[2 e^{i\vec Q \vec \tau}\left(S-\left(n_{k,n}+n_{k,n,\vec{\tau}}\right)\right)\right.\\
&+&\left(e^{i\vec Q \vec \tau}-1\right)\left(a_{-k,n}a_{k,n,\vec{\tau}}e^{i \vec k \vec{\tau}}+a^\dagger_{k,n}a^\dagger_{-k,n,\vec{\tau}}e^{i \vec k \vec{\tau}}\right)+\left.\left(e^{i\vec Q \vec \tau}+1\right)\left(a_{-k,n}a^\dagger_{-k,n,\vec{\tau}}e^{i \vec k \vec{\tau}}+a^\dagger_{k,n}a_{k,n,\vec{\tau}}e^{i \vec k \vec{\tau}}\right)\right]\\
&+&\frac{J_3}{2S}\sum_{k}\sum_{n=1}^{N_u}\sum_{\vec{\tau}_n\neq\vec{\tilde{\tau}}_n} \left[2 e^{i\vec Q (\vec \tau+\vec{\tilde{\tau}})}\left(S-\left(n_{k,n,\vec{\tilde{\tau}}}+2 n_{k,n}+n_{k,n,\vec{\tau}}\right)\right)\right.\\
&+&e^{i\vec Q \vec \tau}\left(e^{i\vec Q \vec{\tilde{\tau}}}-1\right)\left(a_{-k,n}a_{k,n,\vec{\tilde{\tau}}}e^{i \vec k\vec{\tilde{\tau}}}+a^\dagger_{k,n}a^\dagger_{-k,n,\vec{\tilde{\tau}}}e^{i \vec k\vec{\tilde{\tau}}}\right)+e^{i\vec Q \vec \tau}\left(e^{i\vec Q \vec{\tilde{\tau}}}+1\right)\left(a_{-k,n}a^\dagger_{-k,n,\vec{\tilde{\tau}}}e^{i \vec k\vec{\tilde{\tau}}}+a^\dagger_{k,n}a_{k,n,\vec{\tilde{\tau}}}e^{i \vec k \vec{\tilde{\tau}}}\right)\\
&+&e^{i\vec Q \vec{\tilde{\tau}}}\left(e^{i\vec Q \vec \tau}-1\right)\left(a_{-k,n,\vec{\tau}}a_{k,n}e^{-i \vec k \vec{\tau}}+a^\dagger_{k,n,\vec{\tau}}a^\dagger_{-k,n}e^{-i \vec k \vec{\tau}}\right)+\left.e^{i\vec Q \vec{\tilde{\tau}}}\left(e^{i\vec Q \vec \tau}+1\right)\left(a_{-k,n,\vec{\tau}}a^\dagger_{-k,n}e^{-i \vec k \vec{\tau}}+a^\dagger_{k,n,\vec{\tau}}a_{k,n}e^{-i \vec k \vec{\tau}}\right)\right]\\
\end{eqnarray*} 
\end{widetext}
The different phases have different pitch vectors $Q$ and different magnetic cells, leading to different bosonic Hamiltonians.

\subsection{Correction to the energy}
\begin{figure}[t]
\includegraphics[width=0.35\textwidth]{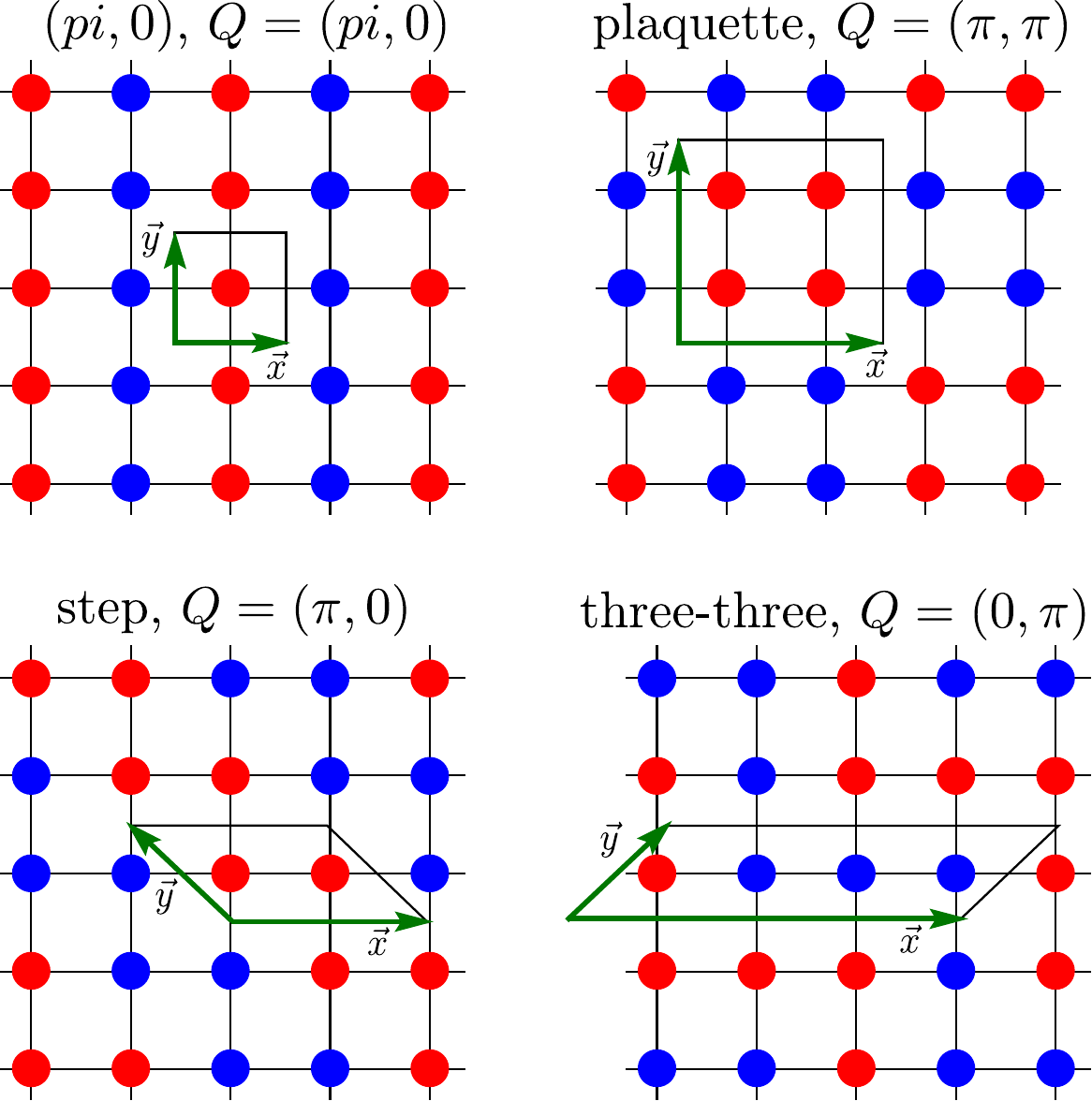}
\caption{The different state around which we compute quantum fluctuations. The elementary cell is depicted in black.}
\label{fig:unit_cell}
\end{figure}

Since for $J_3>1/4$ we have an infinite number of classical ground states, we cannot perform the spin wave calculation for all of them. In particular, we can only treat phases which are periodic. We restrain ourselves to four phases which look paradigmatic to us. The four phases with the associated unit cells are depicted in Fig \ref{fig:unit_cell}. The $(\pi,0)$ phase formed by straight lines of up spins alternating with lines of down spins is the ground state of the $J_1-J_{2^{nd}}$ model for large $J_2/J_1$ and we start our analysis by this phase. We can change this phase by adding domain walls on every other diagonal (which we refer to as the three-three phase), or on every diagonals (which we refer to as the step phase). Finally, the plaquette phase is formed by square of 4 spins pointing all in the same direction alternating with square of spins pointing all in the opposite direction.

\begin{figure}[t]
\includegraphics[width=0.47\textwidth]{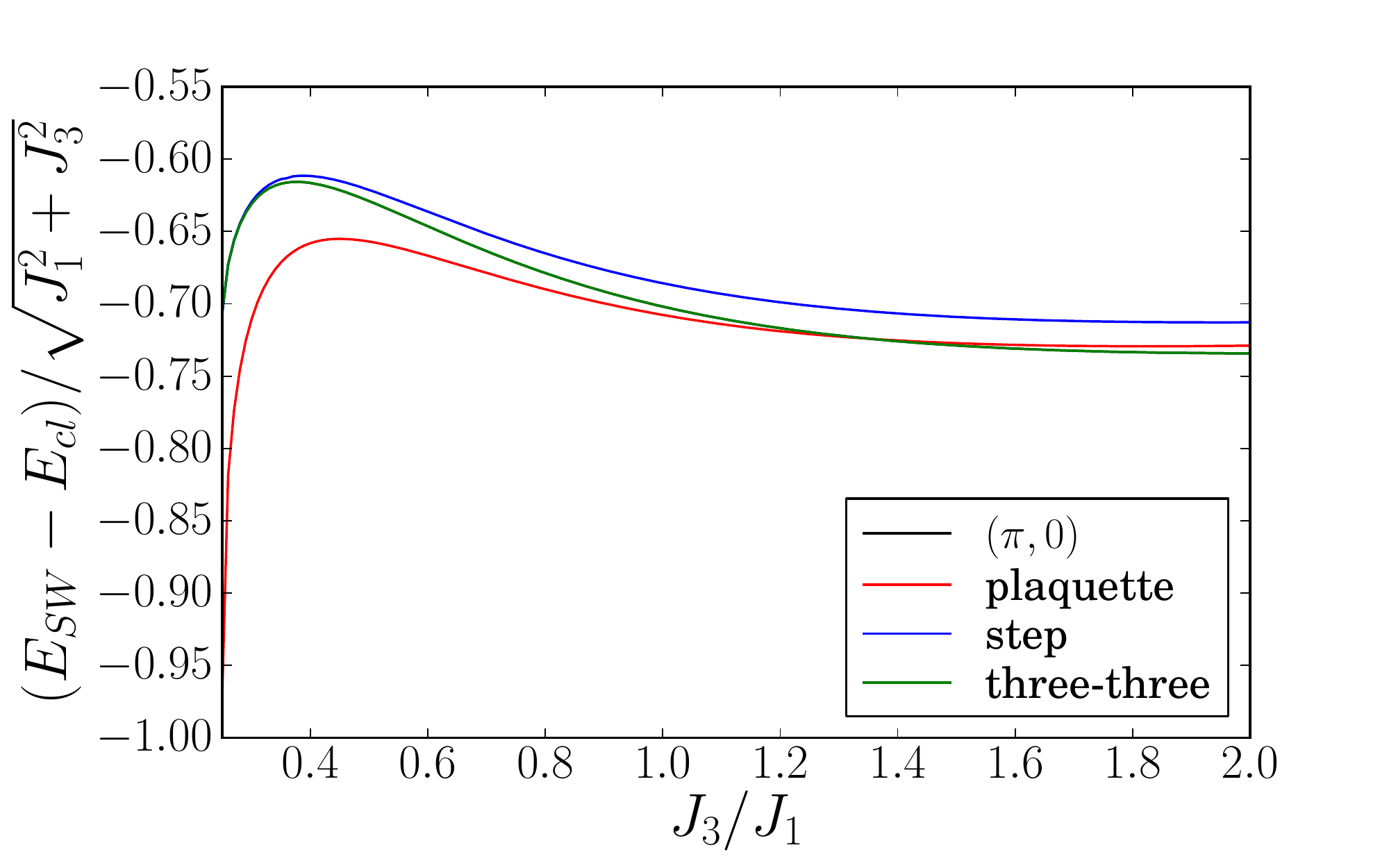}
\caption{Correction to the classical energy due to quantum fluctuations. The $(\pi,0)$ phase and the three-three phase are degenerated. The plaquette phase is selected by quantum fluctuations if $J_3/J_1$ is small enough. For large $J_3/J_1$, the $(\pi,0)$ or the three-three phase are selected. See Fig. \ref{fig:unit_cell} for the description of the phase.}
\label{fig:energy_sw}
\end{figure}

The correction to the energy due to quantum fluctuations is depicted in Fig. \ref{fig:energy_sw}. We see that quantum fluctuations indeed lift the degeneracy between most of the phases already at the $1/S$ level. After the transition from the up-up-down-down phase, quantum fluctuations select the plaquette phase. Around $J_3/J_1\approx 1.37 $, the system undergoes another phase transition due to quantum fluctuations presumably to the $(\pi,0)$ phase. The phase selected after the transition is however not so clear from linear spin wave theory. Indeed, the phase $(\pi,0)$ is degenerate with the three-three phase. We can however speculate that, at higher order, the interaction between the domain walls lift the degeneracy, probably in favor of the phase without any domain wall, considering the fact that the phase with a high density of domain walls is not good energetically. 
\subsection{Reduction of local moment}
It is also useful to consider the average number of bosons to have an estimate of the amplitude of quantum fluctuations. We have computed the fluctuations for all the classical ground states which are present in the spin-wave phase diagram. The helical phases which appear in the mean-field phase diagram are not minimum of the classical energy, and a conventionnal spin-wave theory cannot be performed around them. The calculation of the quantum fluctuations in these cases would require more sophisticated tools which are beyond the scope of this paper.

\begin{figure}[t]
\includegraphics[width=0.47\textwidth]{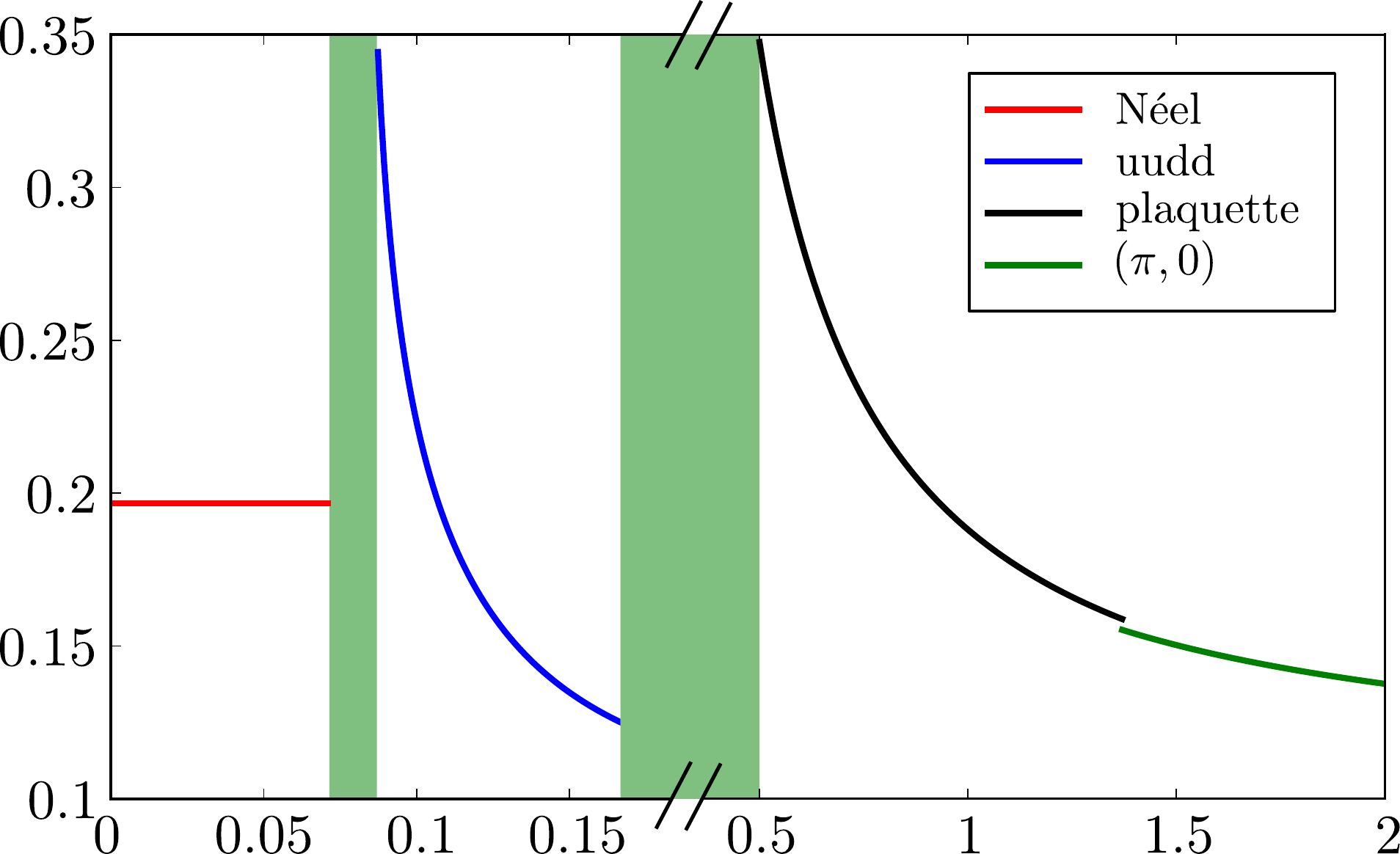}
\caption{Quantum fluctuations computed with linear spin-wave theory for the different classical phases. In green, the region where a helical phase is present and the fluctuations could not be computed.}
\label{fig:fluctuation_sw}
\end{figure}

For a spin-1 system, the projection of the spin along the classical order axis is given by $S = 1-\langle n_{bosons}\rangle$.  If $\langle n_{bosons}\rangle > 2$ the approximation becomes unphysical, and it is well controlled in the limit $\langle n_{bosons}\rangle \ll 1$.
As seen in Fig. \ref{fig:fluctuation_sw}, the fluctuations remain quite small for all four phases, a good indication that the classical ground state is a good approximation of the true ground state in these regions. At $J_3/J_1 = 1/12$ for the $uudd$ phase, and at $J_3/J_1 = 1/4$ for the plaquette phase, the quantum fluctuations diverge (not shown), but within the region where the classical ground states are also minima of the mean-field phase diagram, they remain finite and never exceed $0.35$. 

\begin{figure}[t]
\includegraphics[width=0.48\textwidth]{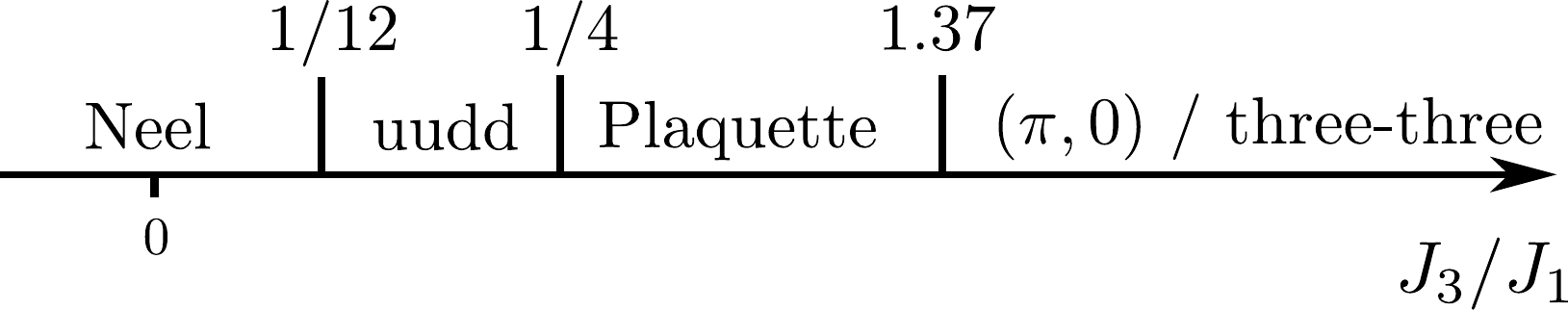}
\caption{Spin wave phase diagram.}
\label{fig:pd_spin_wave}
\end{figure}

\section{Exact diagonalization}
\label{sec:exact-diagonalisation}
\begin{figure}[t]
\includegraphics[width=0.42\textwidth]{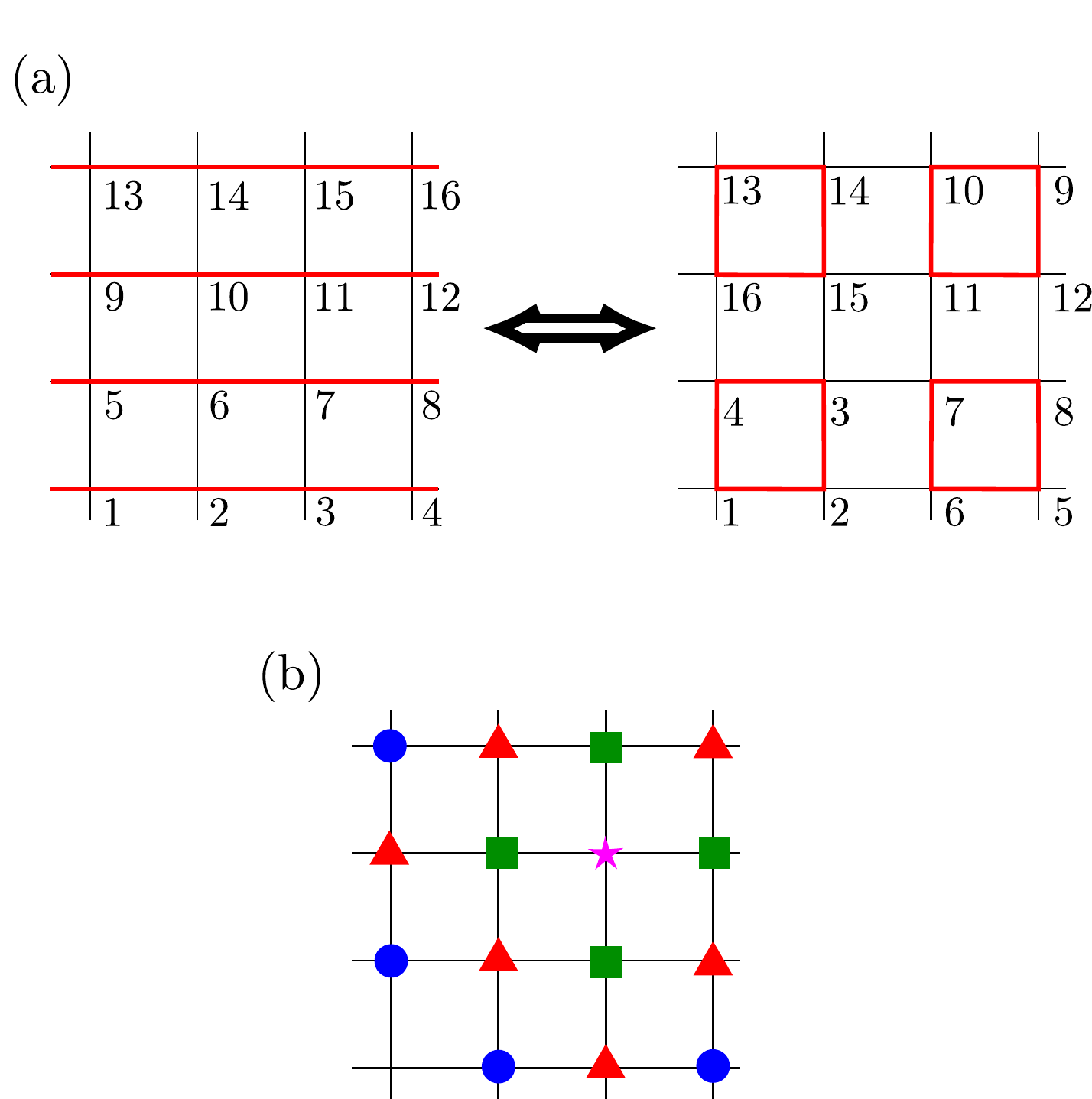}
\caption{(Color online)(a) Hidden symmetries of the square lattice. The Hamiltonian (i.e. the connectivity) is invariant under the operation which exchanges the spin as in the Figure. (b) Class of equivalence of sites with respect to the bottom left site.}
\label{fig:sym_4x4}
\end{figure}

The exact diagonalization approach is very limited for spin $S=1$ system because of the size of the Hilbert space which grows as $3^N$. The only cluster that is compatible with all the classical phases discussed in section \ref{sec:spin-wave} and which is accessible with this method is a $4 \times 4$ cluster. This cluster has some additional symmetries depicted in Fig. \ref{fig:sym_4x4} (a) which make it difficult to treat. As shown in Fig.\ref{fig:sym_4x4} (b), if one picks a reference point, there are only four different types of sites which are not equivalent by symmetry arguments. 

\begin{figure}[t]
\includegraphics[width=0.45\textwidth]{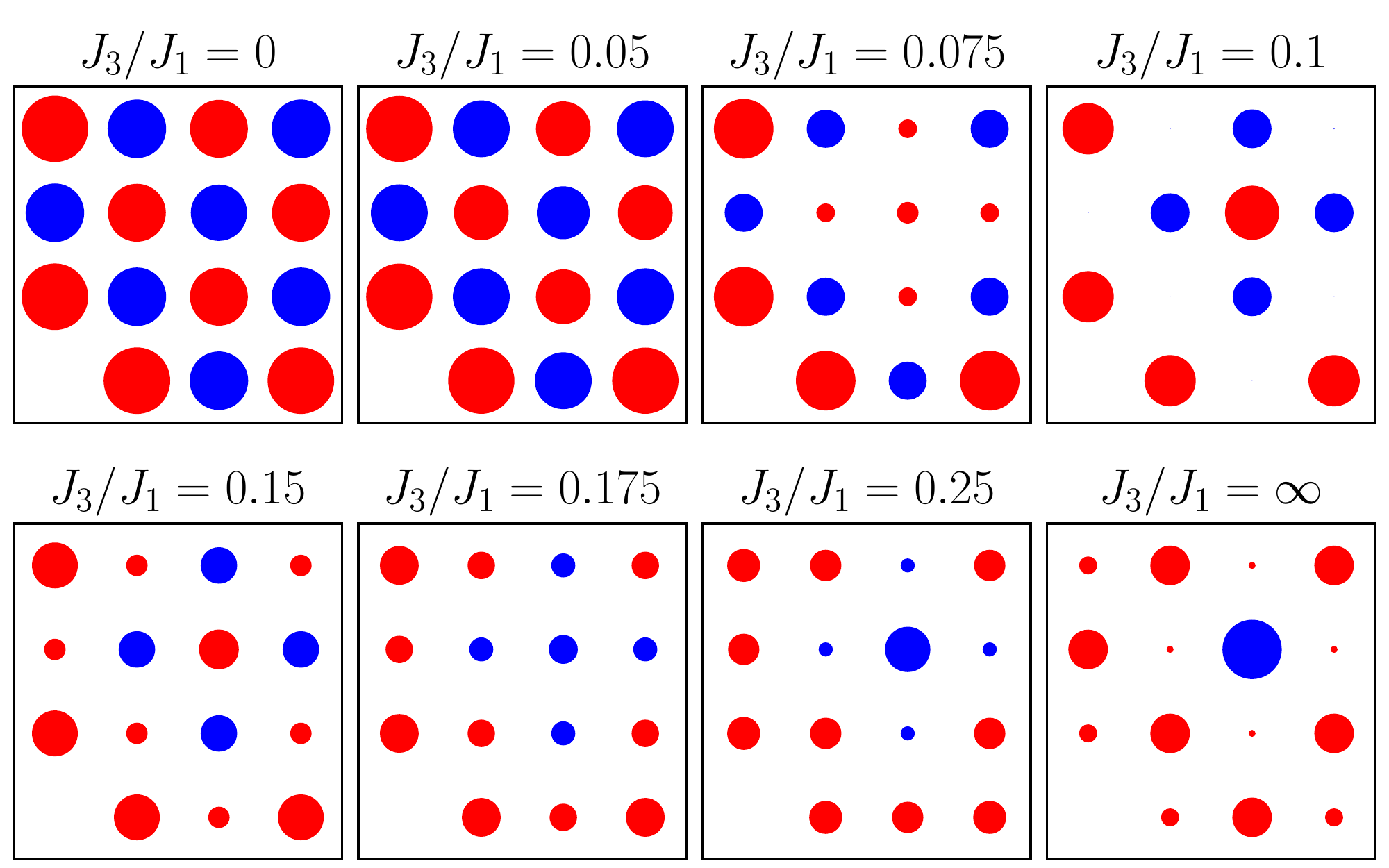}
\caption{(Color online) Spin-Spin correlation with respect to a reference spin at the bottom left for different values of $J_3/J_1$. Red color means a negative correlation while blue means a positive one. The radius of the round is proportional to the absolute value of the correlation.}
\label{fig:ed_corr}
\end{figure}

Despite this fact, we can still check that the obtained results are compatible with exact diagonalization. To do so, we compute the spin-spin correlation function with respect to a reference site. Since we do not expect the quantum ground state to break the symmetry of the Hamiltonian, we have to compare the quantum quantity to a superposition of classical configurations. More precisely, we have to average the quantities over all the classical configurations which are equivalent up to a symmetry of the cluster. In Fig. \ref{fig:ed_corr} we plot the correlation function for different values of $J_3/J_1$. We clearly see at least three different phases. In particular, the second diagonal spin (2,2)\footnote{By this notation, we mean the site at a distance $2 \hat e_x +2 \hat e_y$. }  starts with a ferromagnetic correlation, it then becomes antiferromagnetic and is again ferromagnetic for large $J_3/J_1$. 

\begin{figure*}[t!]
\includegraphics[width=0.98\textwidth]{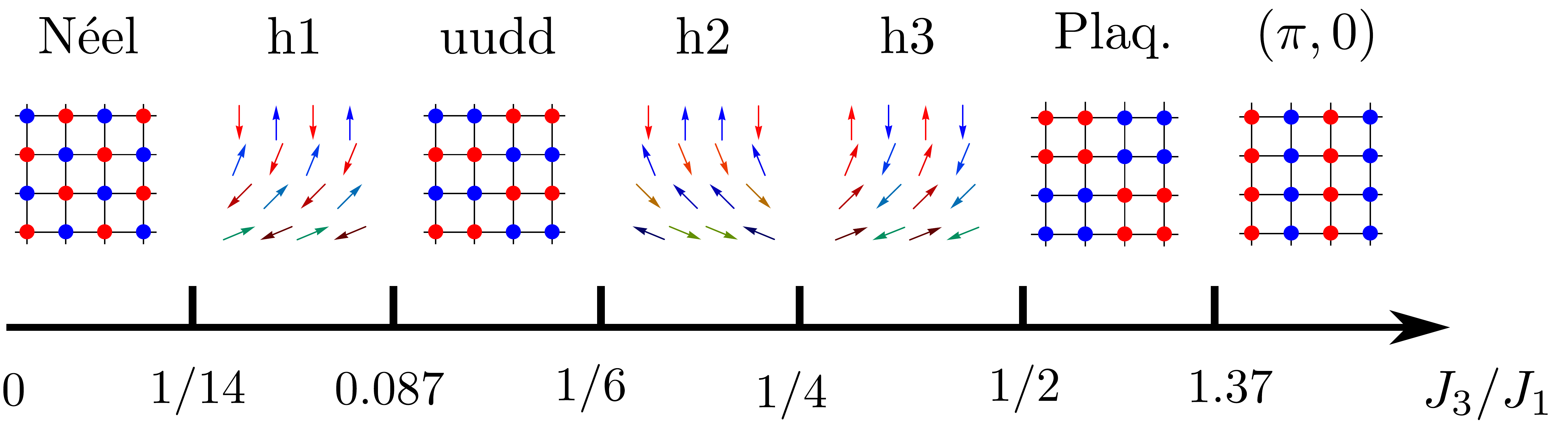}
\caption{Tentative phase diagram of the spin-1 $J_1-J_3$ model on the square lattice.}
\label{fig:pd_final}
\end{figure*}

The value for $J_3=0$ obviously corresponds to a N\'eel state. At $J_3/J_1 = 0.1$, the fifth neighbor (2,2) is strongly antiferromagnetic. This is in agreement with the up-up-down-down phase. Moreover, the second neighbors (1,1) and (2,0) are very close to zero, which is also in agreement with the up-up-down-down phase where half the second neighbors are ferromagnetic and the other half are antiferromagnetic. Finally, the first neighbor is antiferromagnetic, reflecting the fact that in the classical phase, three of them are antiferromagnetic, and one is ferromagnetic.  We are therefore confident that the up-up-down-down phase is stabilized in the thermodynamic limit. 

The exact diagonalization for $J_3/J_1 \gg 1$ does not provide a lot of information. In a $4 \times 4$ cluster, the phase $(\pi,0)$ and the plaquette phase are equivalent by the symmetry arguments presented in Fig. \ref{fig:sym_4x4}. It is therefore hopeless to try to distinguished between these two phases with this cluster. Moreover, the average over the first, second, third and fourth neighbor is the same for all the phases of Fig. \ref{fig:unit_cell}\footnote{This is true only for this cluster. In the thermodynamic limit, for example the second neighbor is different from one phase to the other.}. We therefore have to think about another quantity if we want to make the difference between the plaquette and the $(\pi,0)$ phase and the other phases. One quantity which is different is the product of spins around a loop: $ \left(\vec{S_1}\cdot\vec{S_2} \right)\cdot\left(\vec{S_3}\cdot\vec{S_4}\right)$. This quantity should be positive in the $(\pi,0)$ and in the plaquette phase, and should be negative in the 
step 
phase. It should be negative but small in the case of the three-three phase. We find a value around $0.75$ for $J_3/J_1=\infty$ which clearly points toward the plaquette or the $(\pi,0)$ phase.

\section{Conclusion}
\label{sec:conclusion}
The phase diagram of the spin $S=1$ antiferromagnetic Heisenberg model on the square lattice with three-site interaction turns out
to be extremely rich and complex, with, according to the present results, the possible succession of seven phases, as summarized in Fig. \ref{fig:pd_final}. Already for classical spins the situation is more complicated than for the equivalent $J_{2^{\text{nd}}}-J_{3^{\text{rd}}}$
spin-1/2 model, which has only one transition between a N\'eel phase and a highly degenerate phase, whereas the present model has two
classical phases at small $J_3/J_1$, a N\'eel phase and an up-up-down-down phase, followed by a highly degenerate ground state.
Besides, a mean-field treatment based on a factorized wave-function suggests that the phase diagram is actually more complicated,
with the appearance of three helical phases with a predominantly magnetic order parameter and a tiny quadrupolar component.

Let us now have a critical look at the results. The sequence and stability of the classical phases is in our opinion quite robust: all the phases are stable with respect to semiclassical fluctuations in their range of stability, and the order-by-disorder selection of the plaquette phase for not too large $J_3/J_1$ and of the $(\pi,0)$ phase for large $J_3/J_1$ is plausible in view of the linear-spin wave results, even if, as usual,
the final answer would require to go beyond linear-spin wave theory.

What is less clear however is the fate of the helical phases. Indeed it is well known that, quite generally, quantum fluctuations tend
to favor collinear structures because the harmonic spectrum is softer, and it is not excluded that the helical phases shrink to the
point where they disappear altogether when quantum fluctuations are included, something we have not attempted to do in this paper.
If that turned out to be the case however, the system is likely to develop quantum spin liquid phases at the transition between the N\'eel and the up-up-down-down phases as well as at the transition between the up-up-down-down and the plaquette phases since the correction to the
local moment diverges in the up-up-down-down phase upon approaching 1/12 and in the plaquette phase upon approaching 1/4. This very interesting alternative, as well as the other open issues summarized above, are left for future investigation.

\section{Acknowledgments}
F. Michaud would like to thank T. Coletta for useful discussions about spin waves. The authors acknowledge the Swiss National Fund and MaNEP.

\bibliographystyle{prsty}
\bibliography{bibliography}

\begin{thebibliography}{10}

\bibitem{book_FM}
C. Lacroix, P. Mendels, and F. Mila, {\em Introduction to Frustrated Magnetism:
  Materials, Experiments, Theory}, {\em Springer Series in Solid-State
  Sciences} (Springer, na, 2011).

\bibitem{Dzyaloshinsky}
I. Dzyaloshinsky, Journal of Physics and Chemistry of Solids {\bf 4},  241
  (1958).

\bibitem{Moriya}
T. Moriya, Phys. Rev. {\bf 120},  91  (1960).

\bibitem{Coldea}
R. Coldea {\it et~al.}, Phys. Rev. Lett. {\bf 86},  5377  (2001).

\bibitem{Takahashi}
M. Takahashi, Journal of Physics C: Solid State Physics {\bf 10},  1289
  (1977).

\bibitem{macdonald}
A.~H. MacDonald, S.~M. Girvin, and D. Yoshioka, Phys. Rev. B {\bf 37},  9753
  (1988).

\bibitem{PhysRevB.76.132412}
R. Bastardis, N. Guih\'ery, and C. de~Graaf, Phys. Rev. B {\bf 76},  132412
  (2007).

\bibitem{MVMM}
F. Michaud, F. Vernay, S.~R. Manmana, and F. Mila, Phys. Rev. Lett. {\bf 108},
  127202  (2012).

\bibitem{lai}
C.~K. Lai, Journal of Mathematical Physics {\bf 15},  1675  (1974).

\bibitem{Sutherland}
B. Sutherland, Phys. Rev. B {\bf 12},  3795  (1975).

\bibitem{Takhtajan2}
L.~A. Takhtajan, Physics Letters A {\bf 87},  479   (1982).

\bibitem{Babujian2}
H.~M. Babujian, Nuclear Physics B {\bf 215},  317   (1983).

\bibitem{AKLT}
I. Affleck, T. Kennedy, E.~H. Lieb, and H. Tasaki, Phys. Rev. Lett. {\bf 59},
  799  (1987).

\bibitem{Chubukov}
A.~V. Chubukov, Journal of Physics: Condensed Matter {\bf 2},  1593  (1990).

\bibitem{Fath1}
G. F\'ath and J. S\'olyom, Phys. Rev. B {\bf 44},  11836  (1991).

\bibitem{Fath2}
K. Buchta, G. F\'ath, O. Legeza, and J. S\'olyom, Phys. Rev. B {\bf 72},
  054433  (2005).

\bibitem{Jolicoeur}
U. Schollw\"ock, T. Jolicoeur, and T. Garel, Phys. Rev. B {\bf 53},  3304
  (1996).

\bibitem{Laeuchli}
A. L\"auchli, G. Schmid, and S. Trebst, Phys. Rev. B {\bf 74},  144426  (2006).

\bibitem{Salvatore}
S.~R. Manmana, A.~M. L\"auchli, F.~H.~L. Essler, and F. Mila, Phys. Rev. B {\bf
  83},  184433  (2011).

\bibitem{Kawashima}
K. Harada and N. Kawashima, Phys. Rev. B {\bf 65},  052403  (2002).

\bibitem{BLBQtsunetsugu}
H. Tsunetsugu and M. Arikawa, Journal of the Physical Society of Japan {\bf
  75},  083701  (2006).

\bibitem{BLBQmila}
A. L\"auchli, F. Mila, and K. Penc, Phys. Rev. Lett. {\bf 97},  087205  (2006).

\bibitem{tamas}
T.~A. T\'oth, A.~M. L\"auchli, F. Mila, and K. Penc, Phys. Rev. Lett. {\bf
  105},  265301  (2010).

\bibitem{J1J2transition}
K. Nomura and K. Okamoto, Journal of the Physical Society of Japan {\bf 62},
  1123  (1993).

\bibitem{J1J2J3rastelli}
E. Rastelli, L. Reatto, and A. Tassi, Journal of Physics C: Solid State Physics
  {\bf 19},  6623  (1986).

\bibitem{J1J2J3chandra}
P. Chandra, P. Coleman, and A.~I. Larkin, Journal of Physics: Condensed Matter
  {\bf 2},  7933  (1990).

\bibitem{J1J2J3}
A. Moreo, E. Dagotto, T. Jolicoeur, and J. Riera, Phys. Rev. B {\bf 42},  6283
  (1990).

\bibitem{J1J2J3ferrer}
J. Ferrer, Phys. Rev. B {\bf 47},  8769  (1993).

\bibitem{J1J2J3gochev}
I.~G. Gochev, Phys. Rev. B {\bf 51},  16421  (1995).

\bibitem{J1J2J3yang}
J. Yang, D.-K. Yu, and J.-L. Shen, Physics Letters A {\bf 236},  589   (1997).

\bibitem{J1J2J3m}
M. Mambrini, A. L\"auchli, D. Poilblanc, and F. Mila, Phys. Rev. B {\bf 74},
  144422  (2006).

\bibitem{J1J2J3Arlego}
M. Arlego and W. Brenig, Phys. Rev. B {\bf 78},  224415  (2008).

\bibitem{intro_spin_nematic}
K. Penc and A.~M. Laeuchli,  in {\em Introduction to Frustrated Magnetism},
  Vol.~164 of {\em Springer Series in Solid-State Sciences}, edited by C.
  Lacroix, P. Mendels, and F. Mila (Springer Berlin Heidelberg, na, 2011), pp.\
  331--362.

\bibitem{PhysRev.58.1098}
T. Holstein and H. Primakoff, Phys. Rev. {\bf 58},  1098  (1940).

\end{thebibliography}

\end{document}